\documentclass[%
 reprint,
superscriptaddress,
 amsmath,amssymb,
aps,
prb,
]{revtex4-1}

\usepackage{braket}
\usepackage{graphicx}
\usepackage{dcolumn}
\usepackage{bm}
\usepackage[normalem]{ulem} 
\usepackage{color}

\newcommand{\orcid}[1]{\href{https://orcid.org/#1}{\includegraphics[width=8pt]{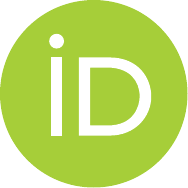}}}

\usepackage{hyperref}
\hypersetup{
    colorlinks,%
    citecolor=blue,%
    linkcolor=blue,%
    urlcolor=blue
}

\begin{document}

\title{Machine learning of microscopic ingredients for graphene oxide/cellulose interaction}

\author{Romana Petry \orcid{0000-0001-9888-4833}}
\affiliation{Brazilian Nanotechnology National Laboratory, CNPEM, C.P. 6192, 13083-970, Campinas, SP, Brazil}
\affiliation{Center for Natural and Human Sciences, Federal University of ABC (UFABC), Santo Andr\'{e}, 09210-580, S\~ao Paulo,Brazil}

\author{Gustavo Silvestre \orcid{0000-0002-4010-2656}}
\affiliation{Instituto de F\'isica, Universidade Federal de Uberl\^andia, C.P. 593, 38400-902 Uberl\^andia, Minas Gerais, Brazil}

\author{Bruno Focassio \orcid{0000-0003-4811-7729}}
\affiliation{Brazilian Nanotechnology National Laboratory, CNPEM, C.P. 6192, 13083-970, Campinas, SP, Brazil}
\affiliation{Center for Natural and Human Sciences, Federal University of ABC (UFABC), Santo Andr\'{e}, 09210-580, S\~ao Paulo,Brazil}

\author{F. Crasto de Lima \orcid{0000-0002-2937-2620}} 
\email{felipe.lima@lnnano.cnpem.br}
\affiliation{Brazilian Nanotechnology National Laboratory, CNPEM, C.P. 6192, 13083-970, Campinas, SP, Brazil}

\author{Roberto H. Miwa \orcid{0000-0002-1237-1525}}
\email{hiroki@ufu.br}
\affiliation{Instituto de F\'isica, Universidade Federal de Uberl\^andia, C.P. 593, 38400-902 Uberl\^andia, Minas Gerais, Brazil}

\author{A. Fazzio \orcid{0000-0001-5384-7676}}
\email{adalberto.fazzio@lnnano.cnpem.br}
\affiliation{Brazilian Nanotechnology National Laboratory, CNPEM, C.P. 6192, 13083-970, Campinas, SP, Brazil}
\affiliation{Center for Natural and Human Sciences, Federal University of ABC (UFABC), Santo Andr\'{e}, 09210-580, S\~ao Paulo,Brazil}

\date{\today}

\begin{abstract}
Understanding the role of microscopic attributes in nanocomposites allows for a controlled and, therefore, acceleration in experimental system designs. In this work, we extracted the relevant parameters controlling the graphene oxide binding strength to cellulose by combining first-principles calculations and machine learning algorithms. We were able to classify the systems among two classes with higher and lower binding energies, which are well defined based on the isolated graphene oxide features. By a theoretical  X-ray photoelectron spectroscopy analysis, we show the extraction of these relevant features. Additionally, we demonstrate the possibilities of a refined control within a machine learning regression between the binding energy values and the system's characteristics. Our work presents a guiding map to the control graphene oxide/cellulose interaction.
\end{abstract}

\maketitle

\section{Introduction}

Aiming for sustainable development, the increasing demand for environmentally friendly materials and renewable resources have led to technological advances in natural polymer applications. For instance, cellulose (C$_6$H$_{10}$O$_5$)$_n$, the most abundant natural polymer on the earth, has been increasingly applied in different fields from food and medicine to nanotechnology. \cite{JIN2021, DONG2021, Trache2020nanocellulose} The nanotechnology applications take part in the physicochemical features presented by nanocelluloses (nCL) while chemical modification and compounding have broadened the scope of its application.\cite{Cel_nanoCS2011}

Nanocelluloses are widely applied in nanocomposite formulations and for the development of advanced hybrid materials. Incorporating different materials (e.g., carbon nanomaterials, metal oxides, polymers) into the cellulosic matrix gives rise to unique properties unrealizable by the components alone. \cite{Trache2020nanocellulose, acs_chemrev_2018} Nanocomposites of nanocelluloses and carbon nanomaterials have been getting attention as a new class of multifunctional hybrid compounds presenting high mechanical strength, electrical and thermal conductivity, and high adsorption capacity.\cite{Zhu_langmuir2019, nanocarbon, nie2020cellulose, Kai2014} In this context, graphene oxide (GO), the oxidized and hydrophilic derivative of graphene, consists in a strategic filler for polymeric composites once it can yield/improve its bioactivity and mechanical/electronic properties.\cite{polymBiocidal, nanocarbon, KAFY2017, GENG2018}

Recently, nCL/GO composites were applied to develop sensors,\cite{SADASIVUNI2015} capacitors,\cite{KAFY2017, SUN2021}  water treatment membranes,\cite{GAO2021, DONG2021} biomaterials,\cite{nanocarbon} and environmental remediation materials.\cite{WANG2021} The GO functional groups (i.e., hydroxyl, epoxy, carbonyl, and carboxyl) favor the electrostatic attraction and hydrogen bonding with cellulose chains.\cite{mao2017interface} Besides, they serve as functionalization and anchoring groups for other materials or molecules.\cite{ELTAWEIL2020, aboamera2018} Furthermore, chemical reduction of GO, resulting in reduced GO (rGO), significantly increases its electronic conductivity by reducing the oxygenated groups and approximating towards the graphene phase.\cite{JPCCganguly2011, CARBONmengying2018} Based on that, rGO has been used to develop conductive cellulosic papers with applications in flexible electrodes.\cite{diao2021sustainable, GAO2013} The wide range of materials, properties, and applications presented by nCL and GO composites is directly related to their structural features and interactions.\cite{mao2017interface} Understanding the role of the functional groups in the GO/nCL interface is essential to guide and refine the composites' functions through a rational materials design. However, the effects of the chemical structure of this interface still need more exploration.

In this work, we used a novel approach to evaluate the chemical and structural relevant parameters governing the binding strength between GO and nCL. Combining first-principles calculations with machine learning automated feature engineering, we extracted the binding energy of different GO/nCL interfaces through density functional theory (DFT) calculations. Here, instead of studying single defects, we take into consideration the surface complexity of GO\cite{RSCADVbouchvakov2013, Motevalli_2019, Defects_ML} by simulating an accurate chemical model on a large supercell with different oxygenated group and oxygen concentrations. We identify two interaction regimes dependent only on the GO features, which we show to be extracted entirely from X-ray photoelectron spectra (XPS). This allowed us to understand the role of microscopic features ruling the GO/nCL interaction strength scale.

\section{Computational Methods}

To generate the data for the latter machine learning classification and regression, we have performed density functional theory calculations as implemented in the VASP package, \cite{CMSkresse1996} with the exchange-correlation term described by the generalized gradient approximation PBE parametrization. \cite{PRLperdew1996} The plane-wave basis was expanded for wave-vectors energies up to $400$\,eV. Given the large supercell, a single k-point was considered for the total energy summation. The electron-ion interaction was described through the projected augmented wave (PAW) method, \cite{PRBblochl1994} with long-range van der Waals interactions described through the vdW-DF functional.\cite{PRLdion2004} The atomic positions were relaxed until the forces in each atom were less than $10^{-2}$\,eV/{\AA}, with a vacuum separation between periodic images of at least $18$\,{\AA}. To find adequate machine learning (ML) descriptors for the classification and regression tasks, we have employed the sure independence screening and sparsifying operator (SISSO) method\cite{PRMouyang2018}. We split the data into train and test with 80/20 ratio. For the classification task, we cross-checked the results of the 80/20 split against a 5-fold cross-validation. For the regression task, we adopted the multi-task method\cite{JPMouyang2019} for high and low binding energies.

\section{Results and discussion}

\subsection{Structure and data generation}

\begin{figure}[h!]
\centering
\includegraphics[width=\columnwidth]{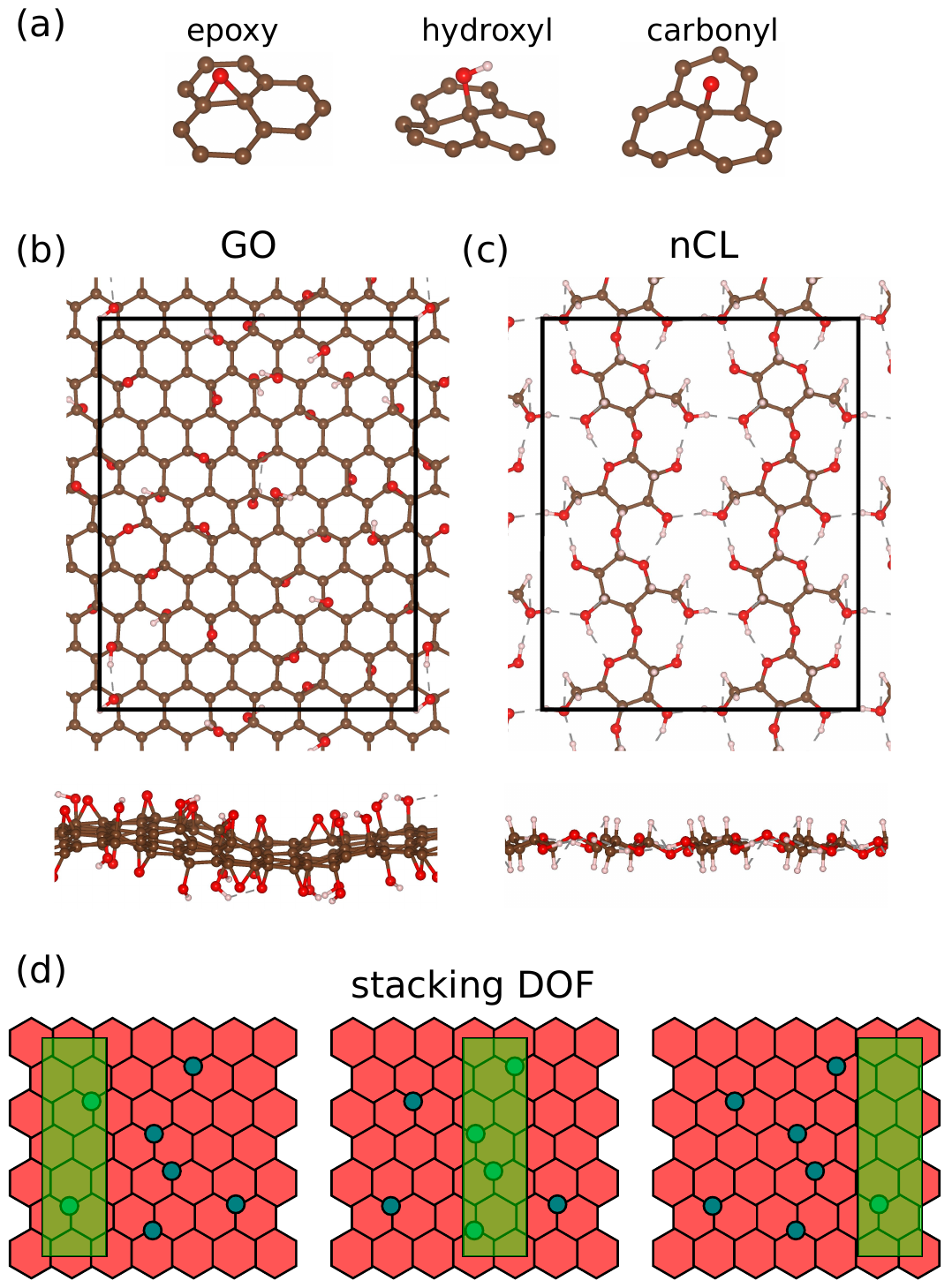}
\caption{\label{fig:str} (a) Oxygenated groups considered on the graphene plane. (b) Graphene oxide and (c) cellulose structural model, the black line mark the considered supercell. (d) Representation of the stacking degree of freedom (DOF) of the cellulose on top of the graphene oxide.}
\end{figure}

The graphene oxide structures were constructed by randomly placing an equal amount of the oxygenated groups epoxy, hydroxyl [Fig.~\ref{fig:str}(a)] on both sides of the graphene sheet. The system was then fully relaxed with the non-bonding atoms (broken during relaxation) removed.\cite{Motevalli_2019, Defects_ML} This final structure has a stoichiometry close to that initially started. It is worth pointing out that carbonyl on the basal plane, resultant from relaxation, can be stabilized by strain fields.\cite{JPCCbagri2010, RSCADVbouchvakov2013} For instance, the corrugation in GO sheets is ruled by the epoxy and hydroxyl observed in our structures [Fig.~\ref{fig:str}(b)]. Following this approach, we have generated 9 GO configurations with oxygen concentration ranging from $6\%$ up to $41\%$; these values are close to those experimentally obtained.\cite{ACSOMliu2021} The GO/nCL interaction was described by a bilayer system composed of a single layer GO lying parallel to a cellulosic sheet.\cite{mao2017interface, xiong2018wrapping} The considered GO supercell [Fig.~\ref{fig:str}(b)] has a lattice mismatch of less than $0.7\%$ from the periodicity of a 2D cellulose cell composed by four cellulosic units (each unit corresponds to two glucose rings) as depicted in Fig.~\ref{fig:str}(c). Since the oxygenated groups are randomly distributed in GO, to get a suitable sampling of the GO/nCL interface energy, we have considered several different interface geometries by performing lateral displacements of the nCL with respect to the GO sheet (for each oxygen concentration) [Fig.~\ref{fig:str}(d)], resulting in a total of 114 different GO/nCL configurations to which we discuss below.

To extract the relevant parameters controlling the binding strength between GO and nCL, we have focus on the features presented in Table~\ref{tab1}. Here, we separate the features into two sets. One depends only on the GO structure before its interaction with the cellulose sheet (pre-interaction). The other ones are extracted after the formation of the GO/nCL interface (post-interaction). The first set of features include the fraction of oxygen ($f_{O}$) and hydrogen ($f_{H}$) with respect to the carbon atoms, which are experimentally controllable during the GO synthesis process. Additionally, we have also considered the density ($n$) of specific types of oxygenated groups in GO, namely hydroxyl ($n_{OH}$), carbonyl ($n_{CO}$), and the sum of epoxy and ether type of groups ($n_{COC}$).

\begin{figure}[h!]
\centering
\includegraphics[width=\columnwidth]{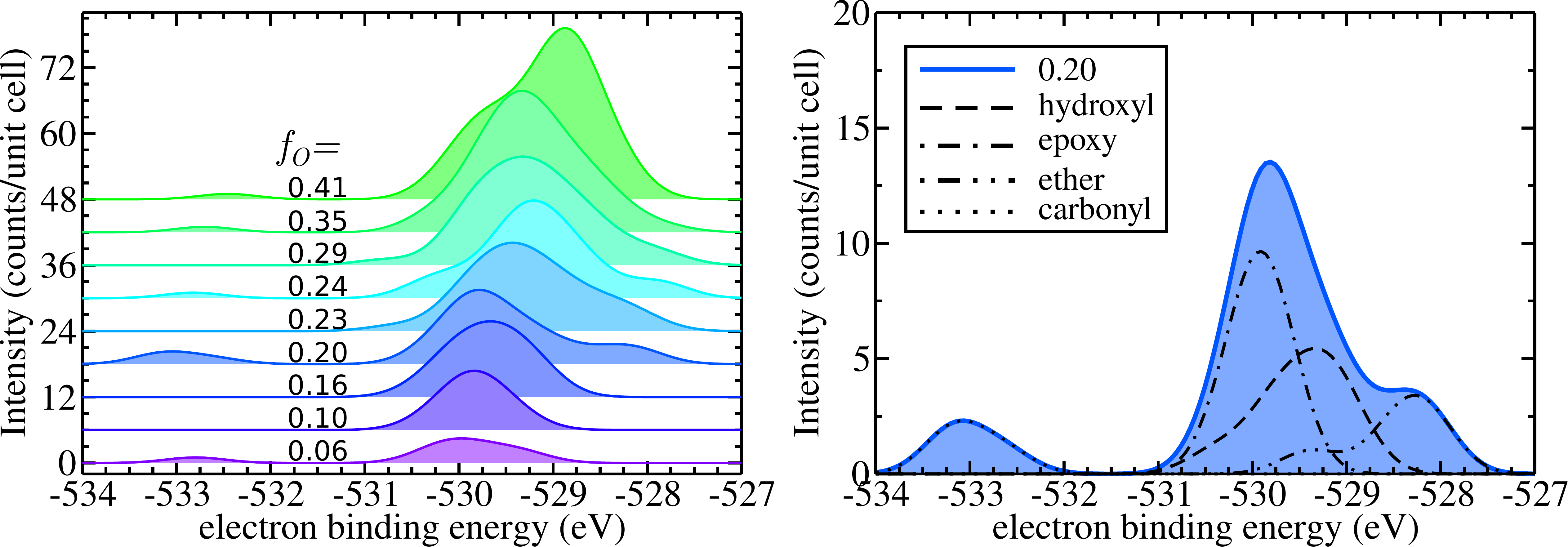}
\caption{\label{fig:xps} Oxygen 1s electron binding energy for (a) all studied GO and (b) detailed analysis for GO with $f_{O}=0.2$. For better visualization a vertical shift of the spectra are presented in (a). We took the C--O characteristic binding energy $-533$\,eV as reference. \cite{JPCCganguly2011, RSCAyu2014}}
\end{figure}

The relative density of these oxygenated groups can be achieved by choosing suitable synthesis routes; while most of structural characterizations have been done through  spectroscopy techniques. Indeed, several works address the atomic structure of GO and GO/nCL interfaces based on XPS analysis.\cite{yao20173d} Here, to provide theoretical support to the experimental identification of the pre-adsorption structure of the GO host, we have performed a set of XPS simulations of the GO's O-$1s$ spectrum as a function of the oxygen fraction ($f_{\rm O}$). As shown in Fig.~\ref{fig:xps}(a), the spectrum intensity is proportional to the oxidation rate, where different functional groups present their characteristic absorption feature. In Fig.~\ref{fig:xps}(b) we present the detailed spectra for the GO with $f_{O}=0.20$. For higher binding energy ($-533$\,eV), a well-defined, individual peak of the carbonyl group is present. Going to lower binding energies, we face respectively the epoxy ($-530$\,eV), hydroxyl ($-529.2$\,eV) and ether ($-528.1$\,eV) peaks. The deconvolution of the whole spectra in those four characteristic peaks allows for determining the relative oxygenated group content on the GO. This by itself permits the determination of the first set of GO's primary feature space (pre-interaction) given in Table~\ref{tab1}. As the major source of hydrogen in the system is through the hydroxyl group, comparing the respective O-1s peak with the C-1s spectra allows the determination of $f_{H}$, and similarly, $f_{O}$, $n_{OH}$, $n_{CO}$ and $n_{COC}$. We will focus on how these features can help classify the GO/nCL interaction strength in the next sections.

The second set of features in Table~\ref{tab1}, post-interaction, dependents on the GO/nCL interface. For instance, we have the vertical corrugation of the carbon atoms of the GO ($\sigma_{GO}$) and the nCL ($\sigma_{nCL}$), and the mean vertical distance ($d_z$) between the GO and nCL carbon atoms layer. Although it can be more difficult to control experimentally, these features can be indirectly manipulated by the features of the first set. In the present study, given the random distribution of the OH groups resulting in hydrogen bonds (HBs) at the GO/nCL interface, we have classified these HBs in three ranges: \cite{CSRemsley1980, PACelangannan2011, PACelangannan2011-2} strong ($n_{Hb1}$) for bond distances smaller than $2.5$\,{\AA}, intermediate ($n_{Hb2}$) for bond distances within $2.5$--$3.2$\,{\AA}, and weak ($n_{Hb3}$) for bond distances within $3.2$--$4.0$\,{\AA}. Additionally, we define the number of carbon $\pi$-stacking bonds ($n_{\pi}$) being the reminiscence of the pristine graphene regions interacting with nCL carbon rings, and a CH$-\pi$ bond ($n_{CH\pi}$) of a hydrogen interacting with an aromatic ring.\cite{CECnishio2004, PCCPnishio2011}

\begin{table}[h!]
\centering
\caption{\label{tab1} Microscopic features of the isolated GO (pre-interaction) and the formed GO/nCL interface (post-interaction). $f_X$ are fraction of $X$ atoms in relation to carbon atoms and are dimensionless; $n_X$ are densities defined per unit cell area ($A_{uc}=3.67\times10^{-14}$cm$^2$); and $d_z$ and $\sigma_X$ are distances in {\AA}.}
\begin{tabular}{ll}
\hline
\hline
\multicolumn{2}{c}{Graphene oxide features} \\
\hline 
Symbol & \multicolumn{1}{c}{Description}  \\
\hline
$f_{O}$   &  fraction of oxygen over carbon atoms \\
$f_{H}$   &  fraction of hydrogen over carbon atoms \\
$n_{OH}$  &  density of hydroxyl type of defects \\
$n_{CO}$  &  density of carbonyl type of defects \\
$n_{COC}$ &  density of epoxy and ether type of defects \\
\hline
\hline
\multicolumn{2}{c}{Interface features} \\
\hline
Symbol & \multicolumn{1}{c}{Description}  \\
\hline
$\sigma_{GO}$ &  GO vertical corrugation \\
$\sigma_{nCL}$ &  nCL vertical corrugation \\
$d_z$         &  mean vertical distance between GO and nCL \\
$n_{Hb1}$     & density of "strong" hydrogen bonds \\
$n_{Hb2}$     & density of "intermediate" hydrogen bonds \\
$n_{Hb3}$     & density of "weak" hydrogen bonds \\
$n_{\pi}$     & density of $\pi$-stacking bonds \\
$n_{CH\pi}$   & density of CH$-\pi$ bonds \\
\hline
\hline
\end{tabular}
\end{table}

\subsection{Binding energy classification}

\begin{figure}[h!]
\centering
\includegraphics[width=\columnwidth]{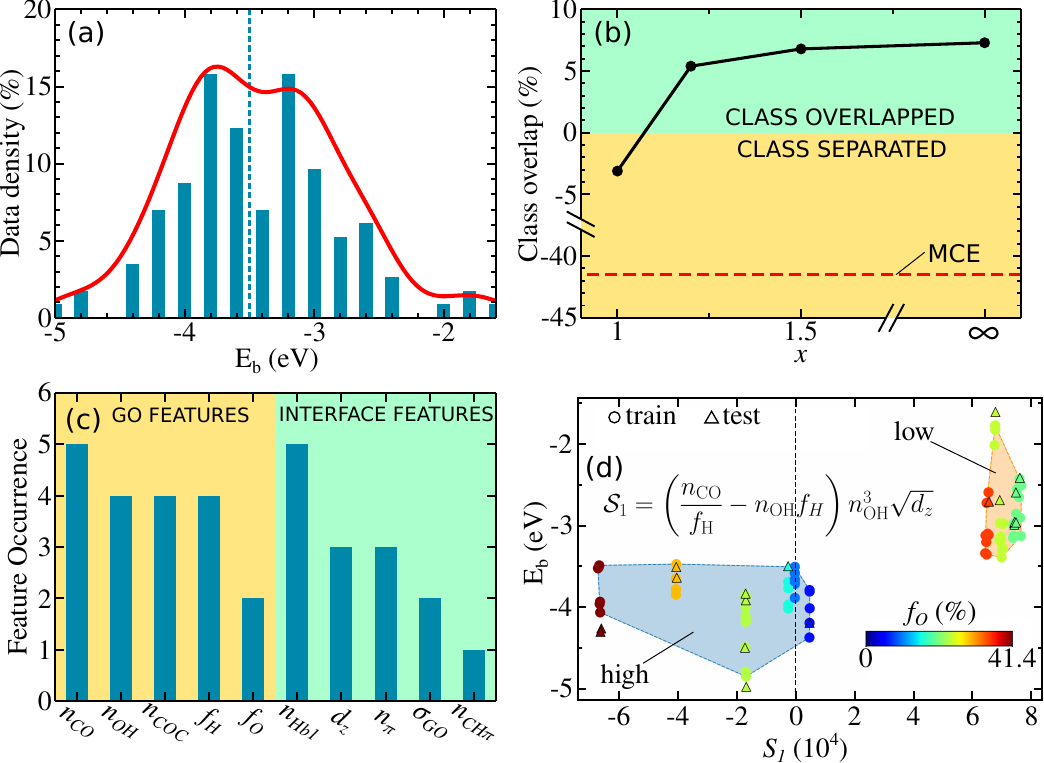}
\caption{\label{fig:class} (a) Distribution of the binding energies ($E_b$) presenting a bimodal behavior. (b) Overlap of classes in different ML models considering different datasets sizes, defined by $x$, negative overlap indicates the classes 1D separation. (c) Feature occurrence in the models presented in panel (b). (d) Visual representation of the class separation in the MCE model.}
\end{figure}

As we have discussed above, the oxygen defects are randomly distributed on the GO layer. Thus the GO/nCL binding strength will depend not only on the concentration of these defects but also on their distribution at the interface region. By analyzing the obtained binding energy for the different stackings of nCL on GO, we see a bimodal behavior [Fig.~\ref{fig:class} (a)]. That gives rise to two possible regimes defining the interface interaction strength. In order to understand and classify a given GO/nCL system within this two regimes, -- i.e., binding energies above/below $-3.5$\,eV per unit cell area ($-10$\,meV/{\AA}$^2$) [indicated by a dashed line on Fig.~\ref{fig:class}(a)] -- we have generated a machine learning classification model.  Here, we have focused on the more interpretative 1D classification model and took into account all the features presented in Table~\ref{tab1}.

On an atomic scale, the GO/nCL binding strength depends on the local geometry at the interface region, which is averaged in an experimental ensemble scenario. In the present study, given the supercell approach, the GO/nCL interface atomic geometry is constrained by periodic boundary conditions, which may result in unrealistic interface configurations. We are aware that the presence of such spurious results can be minimized by increasing the size of the supercell, however, the simulation may become computationally unfeasible. Here, in order to disregard unrealistic GO/nCL interface configurations, however, compromised with the validity of our findings, we have included data points that are within a given range of values close to the mean binding energy ($\langle E_b \rangle$) for a given GO oxidation,
\begin{equation}
\langle {E_b} \rangle - x\,\Delta \le E_b \le \langle {E_b} \rangle + x\,\Delta,
\end{equation}
here $\Delta=\sqrt{\langle E_b^2 \rangle - \langle E_b \rangle^2 }$ is the mean square deviation of the data, and $x$ control the exclusion threshold. Indeed by performing the ML classification for different values of $x$, we see an improvement in the class separation considering a 1D descriptor. We can access this class separation by comparing the value of the class overlap with the spread of the data in the 1D descriptor, where a negative overlap indicates the separation (absolute distance) between classes. For instance, for $x>1.2$ and also considering all generated data ($x=\infty$), a complete separation of two classes was not achieved, being an overlap always present. For $x=1$, a complete separation was possible, where $39$ data points were excluded, with $75$ points composing the data set. Additionally, we performed a rational exclusion removing for each oxide the data belonging to the minority class, namely minority class exclusion (MCE). With this choice, we retain 84 data points to use in the model, more significant than the $x=1$ case and with a better separation between classes [dashed line in Fig.~\ref{fig:class}(b)]. By looking at the features appearing on the descriptors for the different models ($x=1 \rightarrow \infty$ and MCE), we see that the features of the GO alone, pre-interaction features, are more recurrent than those present after the formation of GO/nCL, post-interaction features [Fig.~\ref{fig:class}(c)]. This is particularly evident for the better classifier MCE model, with the descriptor reading
\begin{equation}
S_1 = \left( \frac{n_{CO}}{f_H} - n_{OH} f_H \right) n_{OH}^3 \sqrt{d_z}, \label{eq:s1}
\end{equation}
where we found only $d_z$ as the post-interaction information. As shown in Fig.\,\ref{fig:class}(d), $S_1$ identify the bimodal distribution of the GO/nCL binding energy [Fig.\,\ref{fig:class}(a)]. It is noticeable that the class separation presented can be essentially separated by $S_1<0$ and $S_1>0$ [$S_1=0$ is indicated by a dashed line in Fig.~\ref{fig:class}(d)], with $93\%$ of the data correctly classified in such a way. Moreover, it is interesting to note that the current form of $S_1$  [Eq.~\eqref{eq:s1}] provides the classification of the binding energies and a physical/chemical understanding of the GO/nCL system. For instance, it allow us extract a simple equation ruling the signal of $S_1$, for $S_1<0$
\begin{equation}
\frac{n_{CO}}{n_{OH}} < f_H^2,
\end{equation}
indicating that the increase of the hydroxyl (hydrophilic sites) with respect to the carbonyl groups on the GO surface, will lead to the strengthening of the GO/nCL interaction. Such an increase of the binding energy can be attributed to the energetically more favorable interaction between the hydrophilic-OH groups present in both systems, GO and nCL.\cite{WANG2021} Interestingly, this inequality can be accessed entirely by the XPS spectra as discussed in the previous section. Here, we can propose a focused synthesis of GO to control GO/nCL interaction, with (i) higher hydroxyl density leading to a higher interaction, while (ii) increasing the carbonyl group leads to a lower interaction regime. A reducing process can control the latter, \cite{JPCCganguly2011, CARBONmengying2018} while a decrease in $n_{OH}$ is also achieved.\cite{CAEJchua2013} It is worth pointing out that 5-fold cross-validation of the MCE data, and also a model including all MCE data as training, have resulted in the same 1D descriptor. 

\subsection{Refined binding energy descriptor}

\begin{figure}[h!]
\centering
\includegraphics[width=\columnwidth]{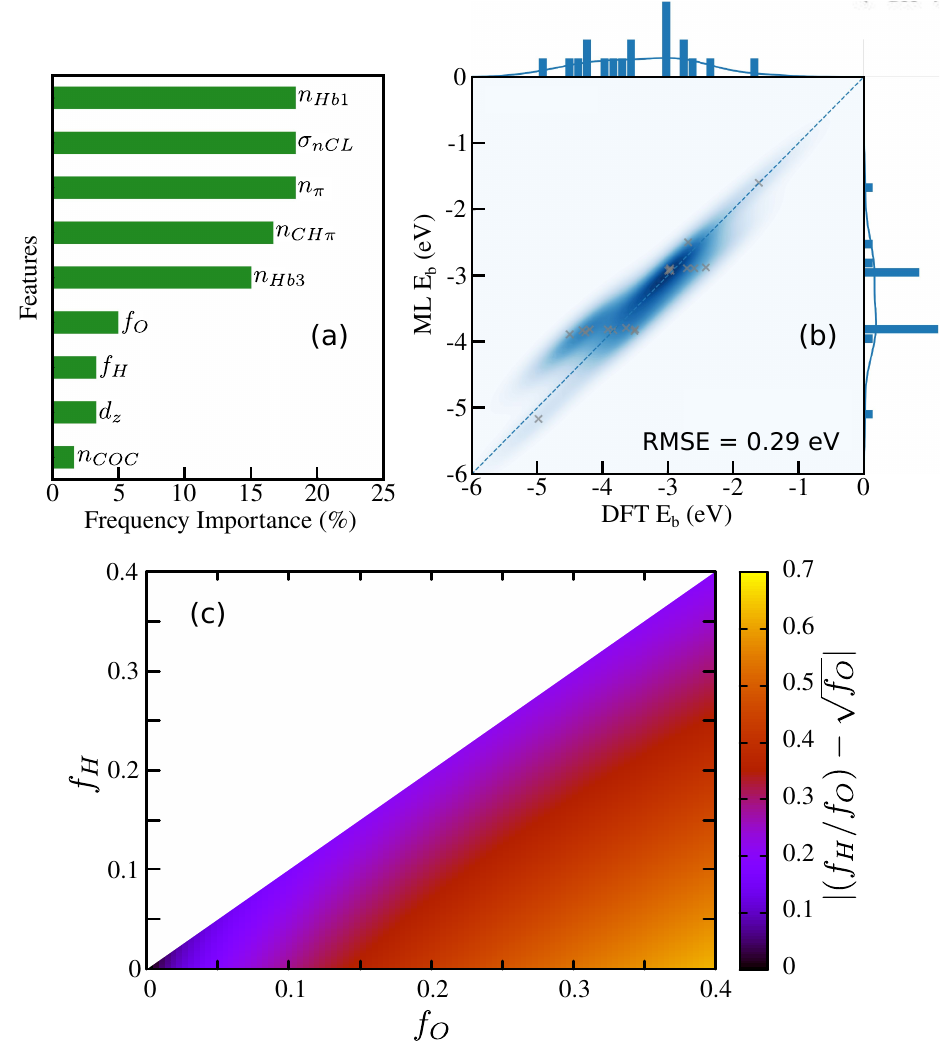}
\caption{\label{fig:regr} (a) Feature occurrence for the 10 1D descriptors with lower root mean square error ($<0.3$\,eV). (b) Predicted binding energy as a function of the calculated one for the test set. (c) Denominator behavior of the best found descriptor in Eq.~\ref{eq:r1}.}
\end{figure}

Going beyond the graphene oxide parameters, the interface features can allow a refined control over the binding energy. Performing a regression within a multi-task approach allow us to extract a 1D descriptor ($R_1$) equivalent for the two classes, to which 
\begin{equation}
    E_b^{ML} = a_i\, R_1 + b_i \label{eq:Eb_ml}
\end{equation}
with $a_i$ and $b_i$, $i=h,l$, representing the higher and lower binding energy class coefficients. Taking the ten most accurate descriptors found by the SISSO algorithm\cite{JPMouyang2019}, all with root mean square error (RMSE) lower than $0.3$\,eV per unit cell ($0.8$\,meV/{\AA}$^2$), we can assess the importance of the features by looking at their occurrence [Fig.~\ref{fig:regr}(a)]. Here the most prominent features on the $E_b$ regression are the strong hydrogen bond ($n_{Hb1}$), the cellulose corrugation ($\sigma_{nCL}$), and the $\pi-\pi$ staking bonds ($n_{\pi}$), followed by the CH$-\pi$ ($n_{CH\pi}$) and weak hydrogen bond ($n_{Hb3}$). In contrast, the oxide pre-interaction features have a lower occurrence.

The better performing descriptor found is
\begin{equation}
    R_1=\frac{n_{CH\pi}^3 - n_{Hb1}^3}{|\left( f_H/f_O \right) - \sqrt{f_O}|}, \label{eq:r1}
\end{equation}
with regression parameters $a_l = 0.24\times10^{-3}$\,eV\,$A_{uc}^3$ and $b_l=-2.87$\,eV for the lower class, while $a_h=0.30\times10^{-4}$\,eV\,$A_{uc}^3$ and $b_h=-3.81$\,eV for the higher binding energy class. The RMSE indicates a fidelity of the predicted $E_b$ value deviating in average by only $8\%$ from the calculated value [Fig.~\ref{fig:regr}(b)]. Note that the $a_i$ coefficients are positive, therefore the difference between $n_{CH\pi}$ and $n_{Hb1}$ rules the GO/nCL binding strength, that is, the sign of $R_1$. This remarkably captures the stronger hydrogen bond ($n_{Hb1}$) energy scale of $0.4$\,eV/bond \cite{PACelangannan2011-2} compared with the weaker CH$-\pi$ interaction $0.05$\,eV/bond;\cite{PCCPnishio2011} and thus in consonance with our previous interpretation of $S_1$. In addition to this competition between the interface H-bond features, the denominator factor is dependent only on the GO features, i.e. pre-interaction features. Although this factor does not change the sign of $R_1$, it modulates the contribution of the interface H-bonds, $n_{Hb1}$ and $n_{CH\pi}$,  dictating the effectiveness of the numerator, $n_{CH\pi}^3 - n_{Hb1}^3$, on the GO/nCL binding energy. Indeed, our results of $|\left( f_H/f_O \right) - \sqrt{f_O}|$ as a function of the $f_H$ and $f_O$, Fig.~\ref{fig:regr}(c), reveal that contribution of   $n_{Hb1}$ and $n_{CH\pi}$ (numerator in $R_1$) becomes more important for $f_O = f_H$. Note that the $f_H$ value is bounded to be $f_H \leq f_O$ with the equality meet for all oxygenated groups being hydroxyl.

Thus, our ML results combined with the oxygenated groups embedded in GO [Fig.\,\ref{fig:str}(a)] and the structural characteristics of the nCL allow us to infer the following interpretation of the GO/nCL binding energy results. As discussed above, (i) the energy gain upon the interaction of the hydroxyl groups with the hydrophilic region of the nCL ($n_{Hb1}$) is more significant compared with that of hydroxyl groups interacting with the hydrophobic region of the nCL ($n_{CH\pi}$) \cite{note}; meanwhile, the epoxy and carbonyl groups interact with the nCL through the formation of (ii) hydrogen bonds,  OH$-$O, at the hydrophilic region of the nCL, and (iii) hydrogen-like bonds, OH$-$C, at the hydrophobic part to the CL. The energy in (ii) is more significant than that in (iii) but both are smaller compared with the one in (i). On the other hand, interactions between both hydrophobic regions are significative for GO/nCL interaction only within the lower oxidized GO, in which the preserved $sp^2$ structure contribute to a significant higher number of $\pi$-stacking interations. Our findings for the ML classification of the GO/nCL binding strength by using the descriptor $S_1$ [(eq.\,(2)],  and further refinement/prediction of binding energy $E^b$ [eq.\,\eqref{eq:Eb_ml}] by the descriptor $R_1$ [eq.\,\eqref{eq:r1}], provide  not only an atomistic understanding of the role played by the oxygenated groups [Fig.\,\ref{fig:str}(a)] on the   GO/nCL binding strength, and but also  a guide for  a refined rational control over the  interface binding energy, for instance, by performing a  selective experimental reduction of carbonyl and epoxy groups.\cite{JPCCganguly2011, CARBONmengying2018}

\section{Conclusions}
In summary, we identify the chemical parameters ruling the energy scale of the graphene oxide binding strength to cellulose. We have generated a set of 114 data of binding energies for different GO oxidation densities employing density functional theory calculations. The observed data allow the inference of two classes with binding energies above and below $10$\,meV/{\AA}$^2$. By an ML classification scheme, we arrive at a descriptor that entirely separates the two classes. The interpretation of such descriptor shows that the GO's different oxygenated group density is the primary attribute ruling the binding energy scale. Performing O1s core level shift calculations, we highlight the possible identification of the higher/lower binding energy regime. Additionally, a multi-task regression of the binding energy over the two GO classes allows us to propose a refined control over the binding energy strength based on the variation of the oxygen density in GO.

\section*{Acknowledgments}

The authors acknowledge financial support from the Brazilian agencies CNPq, FAPEMIG, Capes, FAPESP (grants 18/25103-0, 19/04527-0, 19/20857-0 and 17/02317-2) and INCT-Nanomateriais de Carbono. The authors also acknowledge CENAPAD-SP and Laborat\'orio Nacional de Computa{\c{c}}\~ao Cient\'ifica (SCAFMat-2) for computer time.

%
%


\bibliography{bib}

\begin{thebibliography}{44}
\bibitem [{not()}]{note}%
  \BibitemOpen
  \href@noop {} {\enquote {\bibinfo {title} {{By using the same computational
  approach described in Section 2, we have considered a random distribution of
  OH groups on GO lying near the (i) hydrophilic and (ii) hydrophobic regions
  of the nCL; we found the former configuration [(i)] more stable than the
  latter [(ii)] by about 0.36\,eV/hydroxyl.}}}\ }\BibitemShut {NoStop}%
\end{thebibliography}

%
%
%

\end{document}